\documentclass[prl,longbibliography,showpacs,twocolumn,superscriptaddress,amsmath,amssymb,verbatim]{revtex4-1}

\usepackage{graphicx}
\usepackage{lmodern}
\usepackage{epstopdf}
\usepackage{dcolumn}
\usepackage{bm}
\usepackage{subfigure}
\usepackage{amsmath} 

\usepackage[pdftex,colorlinks=true,citecolor=blue,linkcolor=blue,urlcolor=blue,bookmarks=true]{hyperref}
\usepackage[sort&compress]{natbib}

\usepackage{color}
\usepackage{textcomp}
\definecolor{red}{rgb}{1,0,0}

\definecolor{blue}{rgb}{0,0,1}

\definecolor{green}{rgb}{0,1,0}

\begin{document}
\preprint{APS}

\title{Thermal effects versus spin nematicity in a frustrated spin-1/2 chain}

\author{M. Pregelj}
\email{matej.pregelj@ijs.si}
\affiliation{Jo\v{z}ef Stefan Institute, Jamova c. 39, 1000 Ljubljana, Slovenia}
\author{A. Zorko}
\affiliation{Jo\v{z}ef Stefan Institute, Jamova c. 39, 1000 Ljubljana, Slovenia}
\affiliation{Faculty of Mathematics and Physics, University of Ljubljana, Jadranska u. 19, 1000 Ljubljana, Slovenia}
\author{D. Ar\v{c}on}
\affiliation{Jo\v{z}ef Stefan Institute, Jamova c. 39, 1000 Ljubljana, Slovenia}
\affiliation{Faculty of Mathematics and Physics, University of Ljubljana, Jadranska u. 19, 1000 Ljubljana, Slovenia}
\author{M. Klanj\v{s}ek}
\affiliation{Jo\v{z}ef Stefan Institute, Jamova c. 39, 1000 Ljubljana, Slovenia}
\author{O. Zaharko}
\affiliation{Laboratory for Neutron Scattering and Imaging, PSI, CH-5232 Villigen, Switzerland}
\author{S. Kr\"{a}mer}
\affiliation{Laboratoire National des Champs Magn\'etiques Intenses, LNCMI-CNRS (UPR3228), EMFL, Universit\'e \\ Grenoble Alpes, UPS and INSA Toulouse, Bo\^{i}te Postale 166, 38042 Grenoble Cedex 9, France}
\author{M. Horvati\'{c}}
\affiliation{Laboratoire National des Champs Magn\'etiques Intenses, LNCMI-CNRS (UPR3228), EMFL, Universit\'e \\ Grenoble Alpes, UPS and INSA Toulouse, Bo\^{i}te Postale 166, 38042 Grenoble Cedex 9, France}
\author{A. Prokofiev}
\affiliation{Institute of Solid State Physics, Vienna University of Technology, Wiedner Hauptstrasse 8-10, 1040 Vienna, Austria}

\date{\today}

\begin{abstract}

The spin-nematic phase is an intriguing state of matter that lacks usual long-range dipolar order, yet it exhibits higher multipolar order. 
This makes its detection extremely difficult and controversial.
Recently, nuclear magnetic resonance (NMR) has been proposed as one of the most suitable techniques to confirm its existence.
We report a $^{17}$O NMR observation of the reduction of the local magnetization in the polarized state of the frustrated spin-1/2 chain $\beta$-TeVO$_4$, which was previously proposed to be a fingerprint of the spin-nematic behavior.
However, our detailed study shows that the detected missing fraction of the magnetization, probed by NMR frequency shift, is thermally activated, thus undermining the presence of the spin-nematic phase in the investigated compound.
This highlights the importance of careful considerations of temperature-dependent NMR shift that has been overlooked in previous studies of spin nematicity.

\end{abstract}

\maketitle

The pursuit of a spin-nematic phase \cite{andreev1984spin, lacroix2011introduction} has been attracting physicists for almost half a century \cite{blume1969biquadratic, andreev1984spin, chandra1991quantum, shannon2006nematic, mourigal2012evidence, paddison2015hidden, kohama2019possible, lacroix2011introduction}.
In this enigmatic phase, a spin system develops long-range order of magnetic quadrupoles \cite{andreev1984spin} that lacks dipolar correlations.
Consequently, the spin-nematic phase is elusive and extremely difficult to detect experimentally.
Lately, the research has focused on frustrated spin-1/2 chains, where a spin-nematic phase is predicted to occur in a narrow magnetic-field range just before the magnetization saturates \cite{hikihara2008vector,sudan2009emergent}.
Here, the magnetic quadrupoles are expected to form out of bound magnon pairs that condense at the bonds between the neighboring spins \cite{chubukov1991chiral, shannon2006nematic,zhitomirsky2010magnon}.
The resulting spin-nematic order has been theoretically suggested to be detectable either directly, by probing its excitations, i.e., by breaking bound magnon pairs \cite{sato2009nmr,sato2011field,starykh2014excitations,onishi2015magnetic,furuya2017angular}, or indirectly, by following the corresponding missing fraction of the magnetization \cite{zhitomirsky2010magnon}.
Experimentally, nuclear magnetic resonance (NMR) has been proposed as the most suitable technique to detect the missing fraction due to its high resolution, allowing it to detect minute changes in local magnetization \cite{buttgen2014search, orlova2017nuclear}.
Yet, rounding of the magnetic-field-driven transition into the fully-polarized spin state is characteristic also of thermal fluctuations \cite{kono2015field,blundell2001magnetism}, which have been overlooked in the previous studies of the spin nematicity \cite{buttgen2014search, orlova2017nuclear}.

The most studied frustrated spin-1/2 chain candidate LiCuVO$_4$ has the saturation field between 40 and 50\,T \cite{svistov2011new}.
Its magnetic response in the vicinity of the saturation has been, therefore, investigated primarily by pulsed-field-magnetization and nuclear-magnetic-resonance (NMR) measurements \cite{svistov2011new,buttgen2014search,orlova2017nuclear}.
The latest NMR experiments revealed that at 1.3\,K the magnetic order vanishes already $\sim$1\,T before the full saturation is achieved, which was proposed as an experimental proof of the spin-nematic phase \cite{orlova2017nuclear}.
If the observed response is indeed associated with the emergence of the spin-nematic phase, then it should not change upon cooling and should persists to the lowest accessible temperatures \cite{zhitomirsky2010magnon}, which, however, is yet to be confirmed experimentally.

Here we focus on $\beta$-TeVO$_4$ \cite{meunier1973oxyde,savina2011magnetic}, which also exhibits all the characteristics of the spin-1/2 frustrated ferromagnetic chain \cite{saul2014density,pregelj2015spin,savina2015study,weickert2016magnetic,pregelj2018coexisting}.
The V$^{4+}$ ($S$\,=\,1/2) magnetic ions are coupled by ferromagnetic nearest-neighbor $J_1$\,$\approx$\,$-$38\,K and antiferromagnetic next-nearest-neighbor $J_2$\,$\approx$\,$-J_1$ exchange interactions.
The system exhibits a long-range magnetic order below $T_{N1}$\,=\,4.6\,K and the experimentally derived phase diagram resembles the theoretically predicted one very well \cite{pregelj2015spin}.
In particular, the helical ground state is at $\sim$3\,T succeeded by the spin-density-wave (SDW) state, which is at $\sim$18.7\,T followed by the high-field (HF) phase with an unknown type of incommensurate magnetic order \cite{pregelj2019magnetic}.
Finally, the saturation is reached at $\sim$21.7\,T \cite{pregelj2019magnetic}, allowing for a spin-nematic phase to exist in a narrow interval around this field, as theoretically predicted \cite{singhania2020multiple}.
A relatively low saturation field compared to LiCuVO$_4$ \cite{svistov2011new} and other related systems allows for a much more detailed NMR study of the putative spin-nematic phase, which could focus also on thermal effects.

\begin{figure*}[!]
\centering
\includegraphics[width=\textwidth]{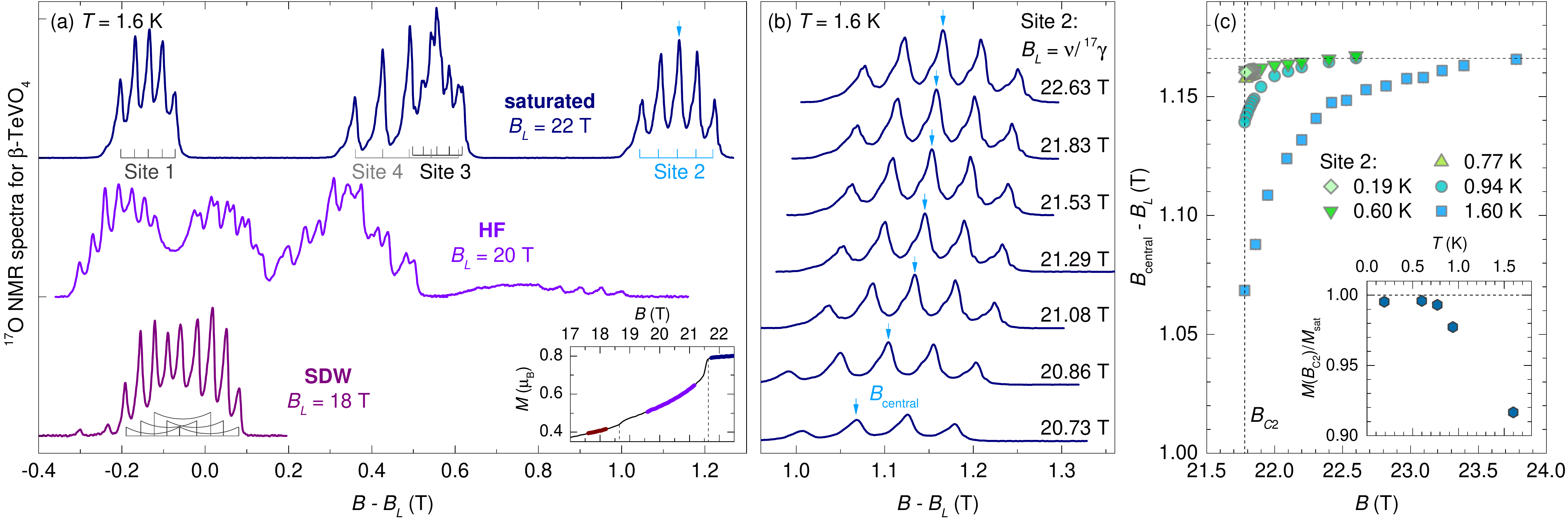}
\caption{The magnetic-field ({\bf B}$||a$) dependence of the $^{17}$O NMR spectra. (a) Spectra measured by sweeping the magnetic field at fixed frequencies $\nu$\,=\,$^{17}\gamma B_L$. The forks at the baseline of the $B_L$\,=\,22\,T curve indicate positions of the spectral peaks (quintets), corresponding to each oxygen site. At the baseline of the $B_L$\,=\,18\,T spectrum U-shaped incommensurate broadening, anticipated for the SDW state, is demonstrated. Inset shows the field dependence of magnetization measured at 1.7\,K \cite{pregelj2019magnetic} with color-codded field ranges covered by NMR spectra. (b) Detailed field dependence of the spectrum for site\,2. Arrows denote the position of the central transition $B_{\text{central}}$. (c) The magnetic field dependence of the shift of the central transitions $B_{\text{central}}$ with respect to $B_L$ for site\,2 measured at several temperatures (symbols). The dashed horizontal line denotes the value ascribed to the fully saturated magnetization $M_{\text{sat}}$. Inset shows the ratio between the magnetization at $B_{C2}$ and $M_{\text{sat}}$ derived from the internal fields at $B_{C2}$ and 23.8\,T, respectively.}
\label{fig-spec}
\end{figure*}

In this work we present low-temperature high-magnetic-field NMR measurements on $\beta$-TeVO$_4$, focusing on the magnetic-field region where the spin-nematic phase could exist.
Our data show that, between 21.72(1) and 21.78(1)\,T, the NMR spectrum progressively changes from a complex one, corresponding to the long-range ordered high-field (HF) phase, to a simple paramagnetic one, indicative of the fully-polarized, i.e., saturated, spin state.
This is characteristic of a first-order transition, where two phases coexist over a finite range of the external magnetic field.
Remarkably, above 21.78\,T the shift of the fully-polarized spectrum relative to the Larmor frequency still grows with increasing field at 1.6\,K, while its width does not change any more.
This resembles the response reported for LiCuVO$_4$ \cite{orlova2017nuclear} and may thus imply the presence of the spin-nematic phase.
However, with decreasing temperature, this shift growth becomes less and less pronounced and is finally completely suppressed below $\sim$0.8\,K.
This reveals that the field-induced shift of the NMR spectrum just below the saturation in $\beta$-TeVO$_4$ is not a sign of the spin-nematic phase, but is rather due to thermal fluctuations.
A precise investigation of the thermal effects on the NMR shift is, therefore, clearly an essential step in the search for the spin-nematic state.
As such, our discovery provides an alternative explanation of the spin-nematic-like behavior that should be considered also in LiCuVO$_4$.

A single crystal (1.00$\times$3.10$\times$7.55\,mm$^3$) of $\beta$-TeVO$_4$ was grown from TeO$_2$ and VO$_2$ powders by chemical vapor transport reaction, using a two-zone furnace and TeCl$_4$ as a transport agent \cite{pregelj2015spin,weickert2016magnetic,pregelj2018coexisting}.
The $^{17}$O enrichment of $\sim$\,8\% was achieved \cite{sup}.
The $^{17}$O NMR was measured using a custom-built spectrometer at the National Laboratory for High Magnetic Fields (LNCMI), Grenoble, France, between 0.2 and 17.7\,K in the magnetic field $B$ between 17 and 24\,T applied along the $a$ crystallographic axis.
A dilution refrigerator was used to achieve temperatures below 1.6\,K.
The spectra were measured at fixed frequencies by sweeping the magnetic field or at fixed fields by sweeping the frequency.
The shape of the coil was adopted to fit the crystal (flat elongated plate), while a rotator was used to adjust the orientation of the magnetic field in the $ab$ plane.
Internal capacitors (bottom tuning) were used to improve signal-to-noise ratio for temperatures above 1.6\,K, while external capacitors (top tuning) were used in combination with the dilution refrigerator.

To explore the high-field magnetic phases in $\beta$-TeVO$_4$ we first measured the $^{17}$O NMR spectrum at 1.6\,K in the SDW, HF and saturated phases, i.e., at $\sim$18, 20 and 22\,T, respectively [Fig.\,\ref{fig-spec}(a)].
For a single $^{17}$O site (nuclear spin $I$\,=\,5/2) one expects a spectrum composed of five lines (a quintet), due to the splitting of the $I$\,=\,5/2 multiplet in an electric field gradient  present in the crystal.
Indeed, in the saturated phase, the spectrum exhibits four quintets, i.e., two well separated and two overlapping quintets, corresponding to four crystallographically inequivalent oxygen sites.
Relative to the ``Larmor'' field $B_L$\,=\,$\nu/^{17}\gamma$, site\,1 is shifted to negative fields by 0.14\,T, whereas sites\,2-4 are shifted to positive fields by 1.14, 0.52, and 0.49\,T, respectively. 
The shift of site\,1 is notably smaller compared to the others, which indicates that the corresponding hyperfine coupling is significantly weaker.
This is the reason why only spectral lines from site\,1 can be detected in the SDW phase, where the signal from other sites is suppressed by fast magnetic relaxation driven by stronger hyperfine interactions.
Since in the SDW phase each of the five NMR transitions is split into a characteristic U-shaped spectrum by incommensurately modulated local magnetic fields, the SDW spectrum consists of five overlapping U-shaped spectral lines [sketched in Fig.\ref{fig-spec}(a)].
In the HF phase, the spectrum is significantly broader and more complex.
In particular, besides the changes of the magnetic order, we find that in this phase also sites 2-4 contribute to the spectrum, indicating that the corresponding magnetic relaxations are slower than in the SDW phase.
The spectrum is completely different than in the saturated phase, implying that the local fields are still continuously distributed, as expected for an incommensurate magnetic order.

In order to investigate the potential spin-nematic behavior, we performed detailed measurements of the NMR spectrum as a function of the applied magnetic field in the vicinity of the magnetization saturation.
Since the local magnetic field is a sum of the applied field and magnetization-driven internal fields, the shifts of the NMR lines relative to $B_L$ continue to change until the magnetization is fully developed. 
This effect is most pronounced for site 2 because it experiences the largest internal field.
At 1.6\,K, we find that between $B_{C2}$\,=\,21.78(1) and 23.4(1)\,T, i.e., where spectrum already adopts a simple saturated shape, the shift of the NMR spectrum of this site relative to $B_L$ still increases by additional 0.1\,T [Fig.\,\ref{fig-spec}(b)].
This becomes even clearer when plotting the shift of the central NMR transition $B_{\text{central}}$ relative to $B_L$ as a function of the applied magnetic field [Fig.\,\ref{fig-spec}(c)] \cite{sup}.
A physical interpretation of this result is that, above $B_{C2}$, the magnetization is still increasing with increasing field.
The observed increase of the ($B_{\text{central}}$\,$-$\,$B_L$) shift above $B_{C2}$ could thus be a signature of the spin-nematic order, as suggested before \cite{orlova2017nuclear}.

To explore the origin of the magnetic-field-induced shift of the fully-polarized spectrum and its potential relation to the spin-nematic order, we performed detailed field-dependent measurements of the central line for site 2 at several temperatures.
The results, summarized in Fig.\,\ref{fig-spec}(c), clearly show that on cooling the corresponding internal field ($B_{\text{central}}$\,$-$\,$B_L$) approaches a field-independent constant, set by the fully-saturated magnetization reached at $\sim$23.8\,T.
Such ``rounding'' of the magnetic-field-driven transition into the fully-polarized spin state, whose sharpness changes with temperature, is obviously due to thermal fluctuations \cite{kono2015field,blundell2001magnetism}.
In fact, since internal fields are in the saturated phase directly proportional to the sample's magnetization, we can plot the ratio between the magnetization at $B_{C2}$ and its fully saturated value [inset in Fig.\,\ref{fig-spec}(c)], which clearly shows that below $\sim$0.6\,K the magnetization is already fully developed at $B_{C2}$.
In contrast, the spin-nematic phase in frustrated spin-1/2 chains is for $J_1/J_2$\,$\sim$\,$-1$ predicted to emerge already at zero temperature when magnetization surpasses $\sim$70\% of its fully saturated value \cite{hikihara2008vector,sudan2009emergent}, i.e., spanning across a sizable magnetization range that should not be suppressed on cooling.
The observed behavior, therefore, indicates that the increase of the internal field above $B_{C2}$ is caused by thermal fluctuations rather than being a signature of the spin-nematic order.

\begin{figure}[!]
\centering
\includegraphics[width=\columnwidth]{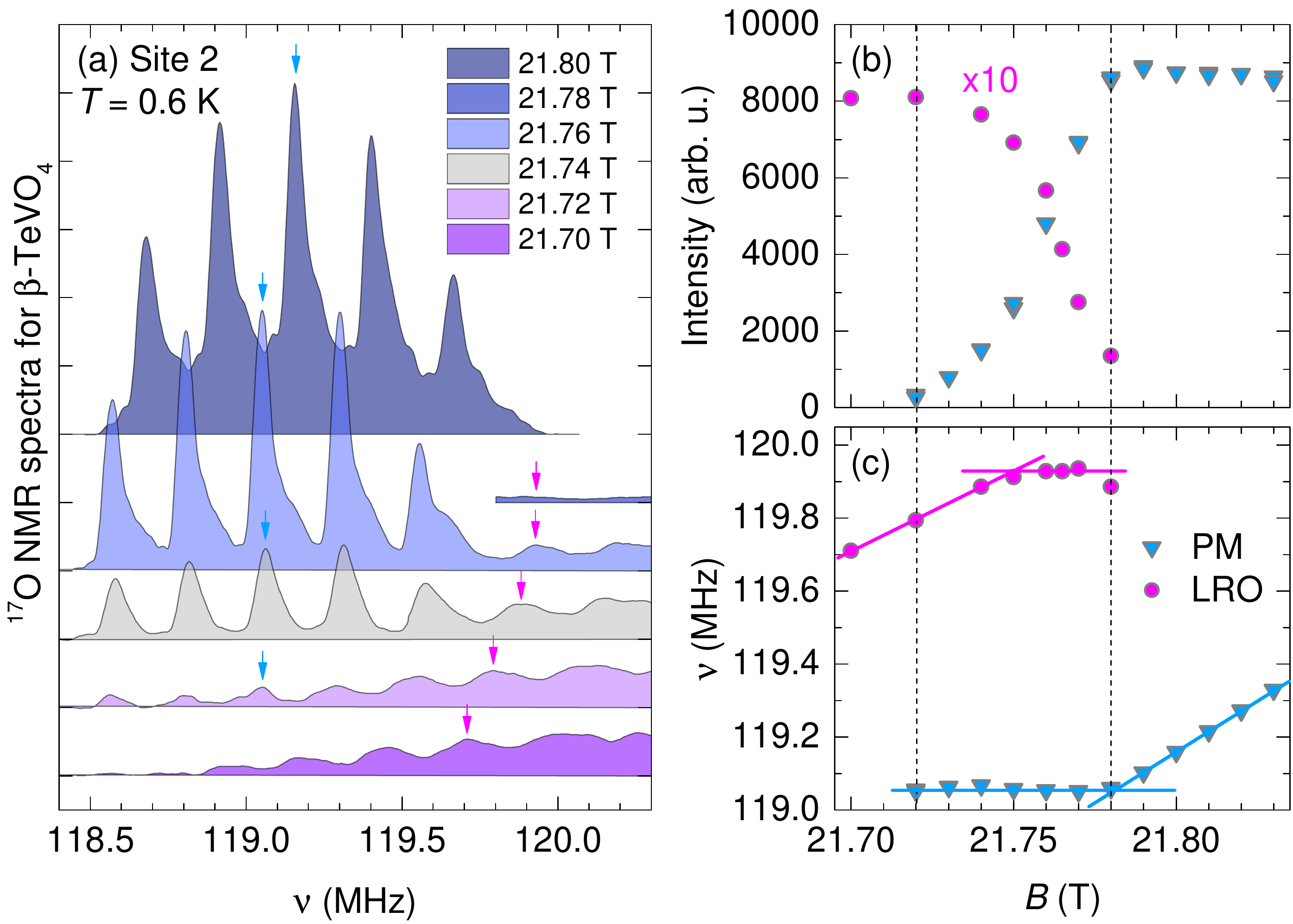}
\caption{(a) Frequency dependent $^{17}$O NMR spectra taken at fixed fields at the transition from the HF to the fully-polarized state measured at 0.6\,K. The field dependencies of (b) intensity and (c) position of the spectral features indicated by arrows in panel (a). Solid lines are guides to the eye.}
\label{fig-mixed}
\end{figure}

A detailed inspection of the transition from the HF state to the saturated state at 0.6\,K reveals another surprising feature.
Namely, across the transition, the NMR spectrum at site\,2 is very complex and corresponds to the HF state coexisting with a simple (paramagnetic-like) fully-polarized, i.e., saturated, state [Fig.\,\ref{fig-mixed}(a)].
On increasing field, the intensity of the fully-polarized spectrum [Fig.\,\ref{fig-mixed}(b)] becomes observable at $B_{C1}$\,=\,21.72(1)\,T and gradually increases up to $B_{C2}$, beyond which field it does not change anymore.
On the contrary, the HF signal starts to loose intensity at $B_{C1}$ and completely disappears at $B_{C2}$ [Fig.\,\ref{fig-mixed}(b)].
Moreover, below $B_{C1}$ and above $B_{C2}$ the positions of the HF and the fully-polarized spectra, respectively, shift linearly with the magnetic field, while throughout the transition they do not change at all [Fig.\,\ref{fig-mixed}(c)].
Such behavior is characteristic of an intermediate mixed-phase state that develops at the first-order transition between two states with different magnetizations \cite{klochan1981spin, bar1972theory,bar1988physics,pregelj2015controllable}, indicating that the transition from the HF state to the saturated state is a first-order type of the phase transition.
In particular, the effect of demagnetizing field makes the local field in the mixed-phase state independent of the applied field \cite{klochan1981spin}, which does not leave any room for a potential spin-nematic phase. 

%

After exploring static properties of the system, we continue with our investigation of spin dynamics by utilizing the nuclear spin-lattice-relaxation ($1/T_1$) measurements.
We focus our attention on the saturated state (above $B_{C2}$), since in frustrated spin-1/2 chains the exotic bound-magnon excitations are expected to be more favorable than conventional single-magnon excitations \cite{zhitomirsky2010magnon}.
We measured $1/T_1$ at the peak of the central transition of site\,2 as a function of magnetic field and temperature.
The data can be nicely fitted with the expression for central transition for $I$\,=\,5/2 nucleus \cite{horvatic1992magnetic} with a stretching exponent of 0.80(3) \cite{sup,mitrovic2008similar}, indicating a finite distribution of $1/T_1$.
The derived dependence of $1/T_1$ [Fig.\,\ref{fig-T1}(a)] exhibits a thermally activated behavior that can be described by the Arrhenius law,
\begin{equation}
1/T_1 = 1/T_1^{\text{HT}}\,\text{exp}(-\Delta/k_BT),
\label{eq-T1}
\end{equation}
where $1/T_1^{\text{HT}}$ is the high-temperature limit for $1/T_1$, $\Delta$ denotes the excitation gap and $k_B$ is the Boltzmann constant.
Indeed, with a field-independent value $1/T_1^{\text{HT}}$\,=\,1.30(1)\,ms$^{-1}$, we obtain a very good agreement with the experimental data in all magnetic fields [solid lines in Fig.\,\ref{fig-T1}(a)].
The derived $\Delta(B)$ is a linear function of $B$  [Fig.\,\ref{fig-T1}(b)]. 
This complies with the theory \cite{zhitomirsky2010magnon, willenberg2016complex}, predicting that at the critical field $B_\Delta$ the bound magnon pairs, constituting the spin-nematic state, give way to the fully saturated state and develop a gap that is proportional to the applied magnetic field. 
Namely, above the saturation the excitation gap should follow the $\Delta$\,=\,$pg\mu_B(B-B_\Delta)$ dependence \cite{buttgen2014search,heinze2019magnetic}, where $g$\,=\,2.01 \cite{pregelj2015spin} is the gyromagnetic ratio and $p$ is the number of bound magnons.
As expected, the field dependence of $\Delta$ can be described perfectly with $B_\Delta$\,=\,19.91(2)\,T and $p$\,=\,1.99(1) [solid line in Fig.\,\ref{fig-T1}(b)], implying that above the saturation the relaxation process is dominated by two-magnon excitations.
For comparison we show a fit assuming one-magnon excitations, which is clearly inadequate [dashed line in Fig.\,\ref{fig-T1}(b)].

However, the derived parameter $B_\Delta$ is smaller than $B_{C1}$, hence, the excitations appear to have a finite gap, $\Delta_{\text{sat}}/k_B$\,$\approx$\,5\,K, already at $B_{C2}$, where the saturated phase is established, in line with the first-order nature of the transition.
Moreover, the derived $\Delta_{\text{sat}}$ is of the order of the exchange anisotropy in the investigated system, $\delta^b_{1,2}/k_B$\,$\sim$\,$|0.2J_{1,2}|/k_B$\,$\sim$7.6\,K \cite{pregelj2018coexisting} and complies with the magnon gap of $\sim$7\,K (0.6\,meV) determined by inelastic neutron scattering (see supplementary information to Ref.\,[\onlinecite{pregelj2019elementary}]).
It thus seems that the exchange anisotropy might be associated with the establishment of the incommensurate HF state instead of the spin-nematic one, as suggested earlier \cite{pregelj2019magnetic}.
This is also in agreement with theoretical calculations, showing that the easy-plane exchange anisotropy and finite interchain interactions, which in $\beta$-TeVO$_4$ amount to $\sim$0.13$|J_1|$ and $\sim$0.05$|J_1|$, respectively \cite{pregelj2018coexisting}, suppress the condensation of bound magnon pairs  \cite{kuzian2007exact,nishimoto2015interplay} and thus preclude the formation of the spin-nematic phase.
Last, we note that a small deviation from the two-magnon behavior is observed in $1/T_1$ at 0.6\,K [Fig.\,\ref{fig-T1}(c)] already below $\sim$21.95\,T, which is far above $B_\Delta$ where the two-magnon spin gap is expected to close, suggesting that it cannot be associated with the spin-nematic fluctuations.

\begin{figure}[!]
\centering
\includegraphics[width=\columnwidth]{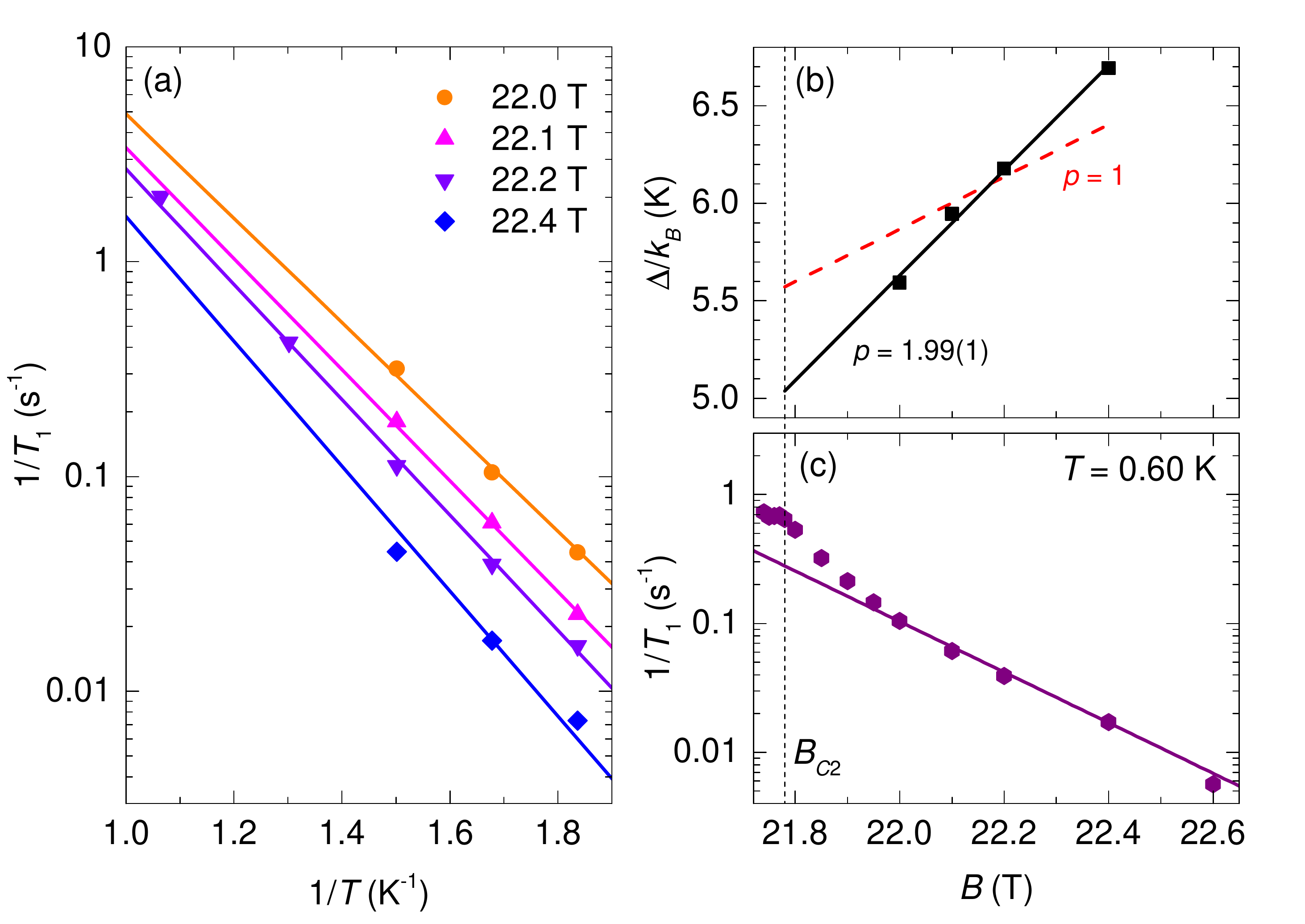}
\caption{(a) The $^{17}$O spin-lattice-relaxation rate measured at several magnetic fields in the saturated phase at the central transition of site\,2 as a function of the inverse temperature. Solid lines are fits to Eq.\,\ref{eq-T1} (see text). (b) The derived excitation gap as a function of the applied magnetic field. Solid line is a linear fit, yielding two magnon excitations ($p$\,=\,2), which is compared to a fit assuming one-magnon exciations ($p$\,=\,1). (c) $1/T_1$ measured at 0.6\,K as a function of the applied magnetic field.}
\label{fig-T1}
\end{figure}
%

Our results particularly highlight the effect of the thermal spin fluctuations.
At finite temperature, the latter push the saturation of magnetization well beyond the zero-temperature limit of 21.78(1)\,T, which can be mistaken for a fingerprint of the spin-nematic order.
In fact, thermal spin fluctuations could also explain the contradictory reports about LiCuVO$_4$, where NMR results for {\bf B}$||c$ and $T$\,=\,0.38\,K \cite{buttgen2014search} show that the magnetization is completely saturated at $\sim$42\,T, whereas the magnetization \cite{buttgen2014search} and NMR \cite{orlova2017nuclear} data obtained at 1.3\,K exhibit a gradual increase of magnetization up to $\sim$44\,T.
The observed behavior was associated either with nonmagnetic defects \cite{buttgen2014search} or with spin-nematic ordering \cite{orlova2017nuclear}, but it may well be a result of thermal spin fluctuations.
Hence, in order to unambiguously confirm the presence of the spin-nematic state in LiCuVO$_4$ an in-depth temperature-dependent investigation at the lowest experimentally accessible  temperatures, like ours, is required.

In conclusion, our high-magnetic-field $^{17}$O NMR measurements on $\beta$-TeVO$_4$ show that below $\sim$1\,K the system undergoes a first-order transition from the HF to the fully-polarized, i.e., saturated, phase at $\sim$21.75\,T.
At lowest temperatures the transition extends across a narrow magnetic field range, which manifests in an intermediate mixed-phase state, where the incommensurate HF and the fully-polarized states coexist.
At elevated temperatures, however, we find that the complete saturation of the magnetization is achieved only at much higher fields, which reflects the effect of thermal fluctuations.
Such a response can easily be mistaken for a spin-nematic behavior potentially undermining the existing reports of spin-nematic phases in other systems.
Our study thus shows that a firm confirmation of the spin-nematic state requires a careful investigation of thermal effects, a crucial experimental step that was not considered in previous studies.

\begin{acknowledgments}

This work has been funded by the Slovenian Research Agency (projects J1-9145 an N1-0148, and program No. P1-0125) and the Swiss National Science Foundation (project SCOPES IZ73Z0\_152734/1).
We acknowledge the support of the LNCMI-CNRS, member of the European Magnetic Field Laboratory (EMFL).

\end{acknowledgments}

\pagebreak
\clearpage
\begin{center}
\textbf{\large Supplemental Material: \\ Thermal effects versus spin nematicity in a frustrated spin-1/2 chain}
\end{center}
\setcounter{equation}{0}
\setcounter{figure}{0}
\setcounter{table}{0}
\setcounter{page}{1}
\makeatletter
\renewcommand{\theequation}{S\arabic{equation}}
\renewcommand{\thefigure}{S\arabic{figure}}
\renewcommand{\bibnumfmt}[1]{[S#1]}
\renewcommand{\citenumfont}[1]{S#1}

\section{Sample preparation}

Starting materials for sample synthesis were TeO$_2$ (99.99\%, Alfa Aesar) and V$_2$O$_5$ (99.9\%, Alfa Aesar).
V$_2$O$_5$ was reduced to V$_2$O$_3$ in a stream of hydrogen-argon mixture (4\% hydrogen) at 750$^\circ$C.
A part of the obtained V$_2$O$_3$ was oxidized by $^{17}$O$_2$ (90\%, Sigma-Aldrich) to V$_3$O$_7$ at 550$^\circ$C.
V$_2$O$_3$, TeO$_2$ and the isotope-rich V$_3$O$_7$ in the stoichiometric ratio providing V$^{4+}$ compound were ground for homogenization, and the fine powder was pressed to a pellet.
This pellet was the starting material for the chemical vapour transport growth of single crystals in a silica ampoule, with TeCl$_4$ as transport agent \cite{pregelj2015spin, pregelj2018coexisting, weickert2016magnetic}.
This technique results in the $^{17}$O enrichment of about 8\% ($\sim$200$\times$ more than the natural abundance).
We expect no variation of the enrichment among them, as all oxygen crystalographic sites are equivalent.

\section{NMR spectrum acquisition}

The nuclear magnetic resonance (NMR) spectra were measured for $^{17}$O nuclei (spin $I$\,=\,5/2) using $\pi/2$\,-\,$\tau$\,-\,$\pi$ pulse sequence with $\tau$ mostly being 30\,$\mu$s, while typical $\pi/2$ and $\pi$ lengths were 3 and 6\,$\mu$s, respectively.
The number of repetitions varied between 32 at 1.6\,K and single shot measurements at 0.19\,K.
The pulse separation was adjusted depending on the spin relaxation rate.
Broad field-sweep measurmemtns, shown in Figs.\,1(a) and 1(b) in the mauscript, were measured at fixed frwquencies while sweeping the magnetic field $B$.
On the other hand, narrow parts of the spectra, shown in Fig.\,2(a) in the manuscript, were meaured at fixed magnetic field while sweeping the excitation frequenciy $\nu$.

\begin{figure}[b]
\centering
\includegraphics[width=\columnwidth]{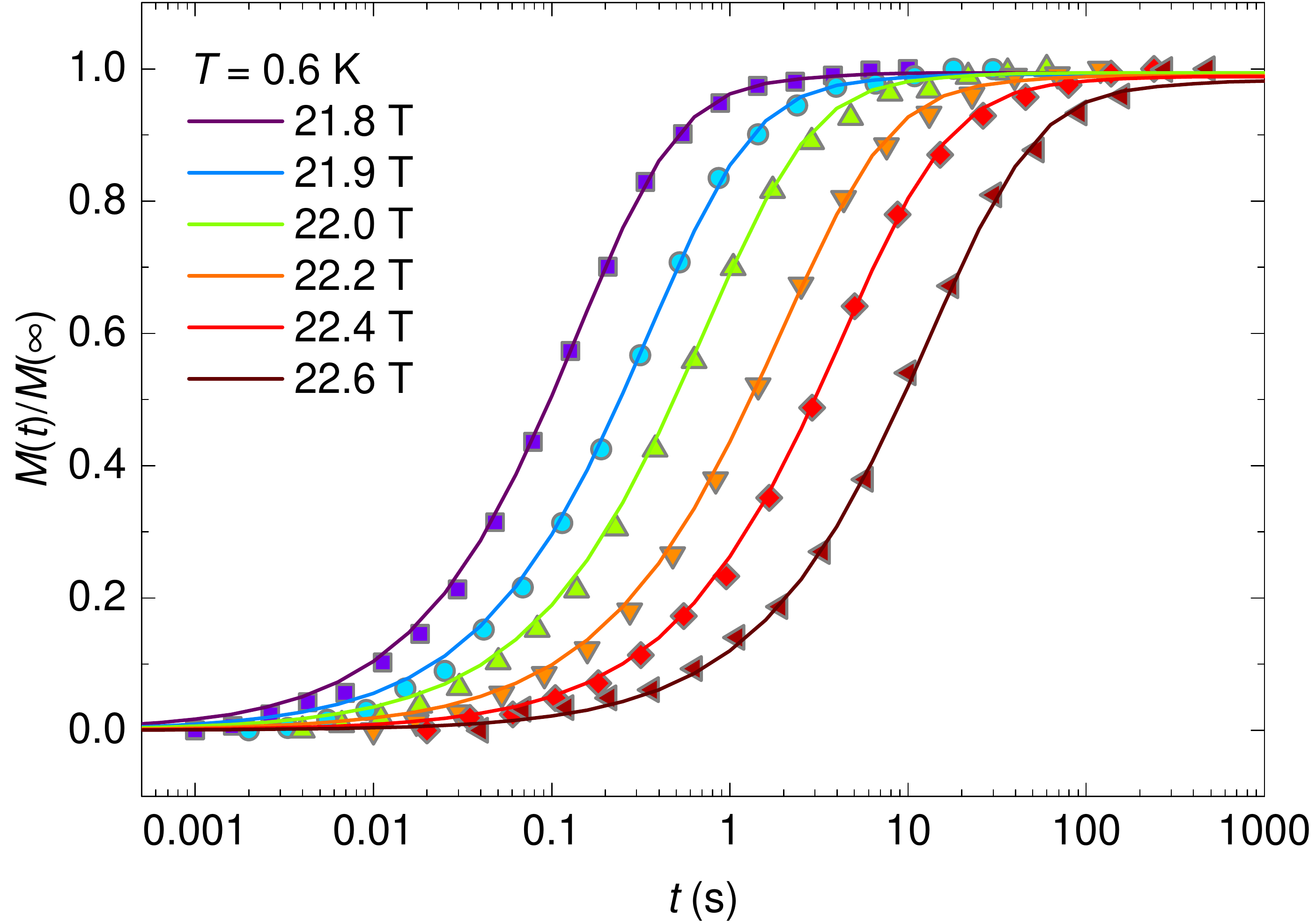}
\caption{Several representative magnetization-recovery curves measured at site\,2 at 0.6\,K. Solid lines are fits to Eq.\,\ref{eq-T1}.}
\label{fig-T1}
\end{figure}


\section{Derivation of $1/T_1$}

The spin-lattice relaxation time $T_1$ was measured by inversion-recovery method at the central line of the NMR spectrum. 
The magnetization recovery curves were fitted by the corresponding magnetic-relaxation model for spin-5/2 nuclei \cite{horvatic1992magnetic},

\begin{equation}
\begin{split}
M(t) =   M_0\bigg[1- & \left( \frac{1}{35}\text{e}^{-(t/T_1)^\beta}+\frac{8}{45}\text{e}^{-(6t/T_1)^\beta}+ \right.  \\
& \quad \left. \frac{50}{63}\text{e}^{-(15t/T_1)^\beta}\right)\bigg].
\label{eq-T1}
\end{split}
\end{equation}
Here $M_0$ is the saturation magnetization and $\beta$ denotes a stretched exponential relaxation, which takes into the account the distribution of $T_1$ values.
Eq.\,\ref{eq-T1} with $\beta$\,=\,0.80(3) perfectly fits the experimental curves for all magnetic fields between 21.7 and 23.8\,T (Fig.\,\ref{fig-T1}).
The derived values for $1/T_1$ are presented in Fig.\,3 in the main text.

\end{document}